\documentclass[twocolumn]{aa}
\usepackage{graphicx}
\usepackage[authoryear]{natbib}
\bibpunct{(}{)}{;}{a}{}{,}
\usepackage{txfonts}
\begin{document}

\title{Large-scale study of the NGC\,1399 globular cluster system in
Fornax}

\author{L. P. Bassino\inst{1,2}\thanks{
E-mails:\,lbassino@fcaglp.unlp.edu.ar\,(LPB); 
favio@fcaglp.unlp.edu.ar\,(FRF);\,forte@fcaglp.unlp.edu.ar\,(JCF); 
bdirsch@cepheid.cfm.udec.cl(BD); 
tom@mobydick.cfm.udec.cl\,(TR);  
dgeisler@astro-udec.cl\,(DG); 
ylva@astro.uni-bonn.de\,(YS)}
          \and
          F. R. Faifer\inst{1,2}$^{\star}$
          \and
           J. C. Forte\inst{1}$^{\star}$
          \and
           B. Dirsch\inst{3}$^{\star}$ 
           \and 
           T. Richtler\inst{3}$^{\star}$
           \and  
           D. Geisler\inst{3}$^{\star}$
           \and 
           Y. Schuberth\inst{4}$^{\star}$
          }

\institute{CONICET and Facultad de Ciencias Astron\'omicas y Geof\'{\i}sicas,
       Universidad Nacional de La Plata, Paseo del Bosque S/N, 1900-La Plata,
       Argentina
\and
IALP - CONICET, Argentina
\and
Universidad de Concepci\'on, Departamento de F\'{\i}sica,
      Casilla 160-C, Concepci\'on, Chile
\and
Sternwarte der Universit\"at Bonn, Auf dem H\"ugel 71, D-53121 Bonn, 
Germany}      
 
\offprints{L. P. Bassino}
\date{Received  / Accepted  }

\titlerunning{Large-scale study of the NGC\,1399 globular cluster system}
\authorrunning{Bassino et al.}

\abstract{
We present a Washington $C$ and Kron--Cousins $R$ photometric study of the 
globular cluster system of NGC\,1399, the central galaxy of the Fornax cluster.
A large areal coverage of 1~square degree around NGC\,1399 is achieved with 
three adjoining fields of the MOSAIC\,II Imager at the CTIO 4-m telescope.
Working on such a large field, we can perform  
the first indicative determination of the total size of the NGC\,1399 
globular cluster system. The estimated angular extent, measured from 
the NGC\,1399 centre and up to a limiting radius where the areal density of 
blue globular clusters falls to 30 per cent of the background level, 
is 45 $\pm$ 5 arcmin, 
which corresponds to 220~--~275 kpc at the Fornax distance. 
The bimodal colour distribution of this globular cluster system,  
as well as the different radial distribution of blue and red clusters, up 
to these large distances from the parent galaxy, are confirmed. 
The azimuthal globular cluster distribution exhibits asymmetries that 
might be understood in terms of tidal stripping of globulars from 
NGC\,1387, a nearby galaxy. The good agreement between the areal 
density profile of blue clusters and a projected dark-matter NFW 
density profile is emphasized. 

}

 \maketitle     
\section{Introduction}

The study of globular cluster systems (GCSs) around giant galaxies   
within galaxy clusters has proved to be a useful tool for improving 
our knowledge of the formation and evolution of galaxies in 
these environments, e.g. M\,87 \citep{han01,cot01} and M\,49 
\citep{rho01,cot03} in Virgo, NGC\,1399 in Fornax \citep{ost98,dir03} or 
NGC\,3311 in Hydra \citep{mcl95,bro00}.  
Many of these GCSs are so huge that it has not been possible yet to study 
most of the system through their full projected spatial distribution even 
with wide-field mosaics \citep{rho01,dir03}.  

In particular, much work has been devoted to the GCS around NGC\,1399, at 
the centre of the Fornax cluster. The wide-field study by \citet[ hereafter 
Paper I]{dir03} presents a detailed description of the previous investigations 
of this GCS so we simply refer the reader to it instead of repeating it 
here. The Washington 
photometry presented in Paper\,I is based on images from the CTIO MOSAIC camera 
($36 \times 36$ arcmin) obtained with the Washington photometric system. According 
to this study the GCS extends farther than 100 kpc from the parent galaxy 
(NGC\,1399). The globular cluster (GC) colour distribution is bimodal, and 
for r $<$ 8 arcmin the (projected) spatial distribution of the red (
metal--rich) GCs is more concentrated than that of the blue (metal--poor) 
clusters. At larger galactocentric distances, however, the radial density 
profiles are similar. The total number of GCs is estimated to be about 6500, 
and the global specific frequency $S_{N}$ \citep[as defined by][]{har81}  
is found to be $5.1 \pm 1.2$. A further discussion of the 
intrinsic specific frequencies of both cluster populations, as well as of the
connection of the colour bimodality with the surface profile brightness of
the galaxy, has been presented in \citet{for05}. 
    
The dynamical status of the GCS around NGC\,1399 has been addressed in  
recent kinematic studies by \citet{ric04,dir04,sch04}. Another wide-field 
study of the GCS of NGC\,4636, a bright elliptical in the 
Virgo cluster, has recently been performed by \citet{dir05} on the basis 
of the same type of data as used in the present study: wide-field Washington 
photometry. 

Here we present a new study of the NGC\,1399 GCS based on an  
enlarged set of MOSAIC wide-field images. In addition to the  
`central field' analysed in Paper~I, we obtained Washington photometry of two 
new adjoining fields labelled `east' and `northeast'. The larger area  
now covered enables us to assess the approximate extent of the GCS and to 
perform a large-scale study of the colour and spatial distribution of 
the blue and red cluster populations.

This analysis of the NGC\,1399 GCS is organised as follows.  
Section 2 describes the observations and GC candidate selection. 
The colour and spatial distributions are analysed in Sec. 3. 
Finally, a summary and a discussion of the results are 
presented in Sec. 4.

\section{Observations and reductions}

The images used in this study were obtained during three observing runs  
with the MOSAIC camera (8 CCDs mosaic imager) mounted at the prime  
focus of the 4-m Blanco telescope at the Cerro Tololo Inter--American 
Observatory (CTIO). The images containing NGC\,1399 (`central field') and 
the background field located 3.5 deg northeast of this galaxy have 
already been presented in Paper~I which we refer the reader to for 
details of the observations. 
Observations of the two fields adjacent to NGC\,1399 (`east' and 
`northeast' fields)  were carried out during November 17--19, 2001. 
All fields (except for the background) overlap each other and are 
shown in Fig.~1. The wide field of the MOSAIC is 36 $\times$ 36 arcmin 
(200 $\times$ 200 kpc at the Fornax distance) with a pixel scale of 
0.27 arcsec. In the following, the adopted distance to 
NGC\,1399 is $m-M = 31.4$ mag (Paper~I) corresponding to 19\,Mpc. 

All fields were imaged in Kron--Cousins $R$ and Washington $C$. 
We selected the $R$ filter instead of the original Washington $T_1$,  
as the Kron--Cousins $R$ and $T_1$ magnitudes are very similar (only a 
zero-point difference $R - T_1 \approx -0.02$) and the $R$ filter is more 
efficient than $T_1$ \citep{gei96}. As in the previously observed fields,  
to fill in the gaps between the 8 individual MOSAIC chips, the data 
in the two adjacent fields were dithered taking three images in $R$ with 
exposure times of 600\,s each, and five images for the `east' field and 
four for the `northeast' one, in $C$, with exposures of 1\,200\,s each. 

\begin{figure}
\resizebox{\hsize}{!}{\includegraphics{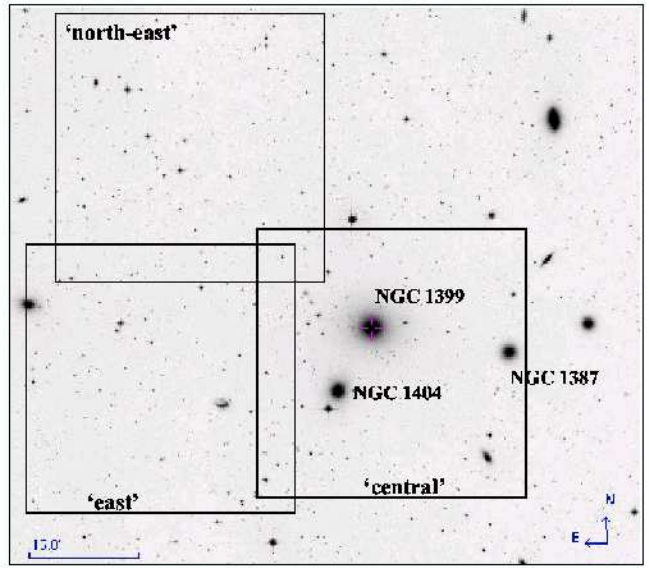}}
\caption{MOSAIC fields, labelled `central', `east', and `northeast', overlaid 
on a DSS image of the central region of the Fornax cluster. North is up and 
East to the left.}
\end{figure}

\begin{figure}
\resizebox{\hsize}{!}{\includegraphics{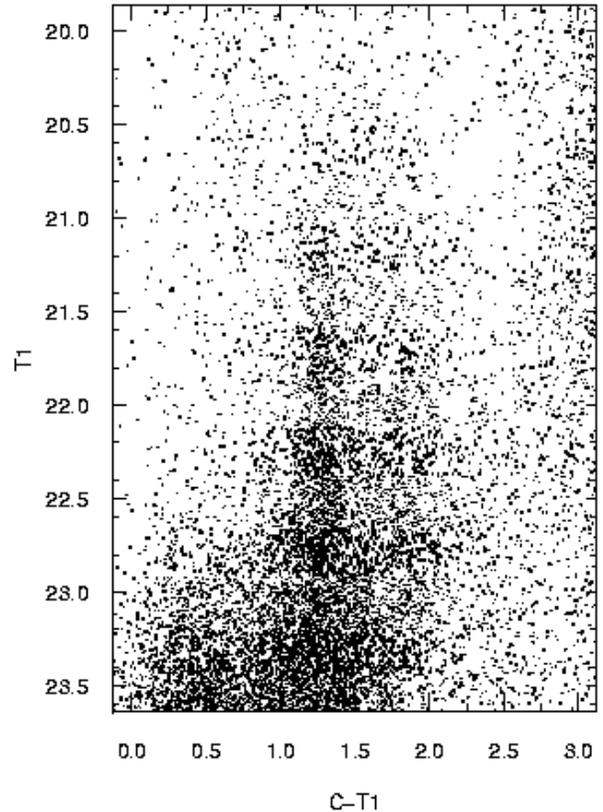}}
\caption{Colour-magnitude diagram for all point sources in the MOSAIC 
field. Globular cluster candidates show up in the colour range 
$0.8 \leq (C-T_1) \leq 2.3$. The dense group of objects bluer than 
$(C-T_1) = 0.8$ and fainter than $ T_1 = 23$ are mainly compact 
background galaxies.}
\end{figure}

The MOSAIC data in the `east' and `northeast' fields were reduced 
using the {\it mscred} package within IRAF\footnote{IRAF is distributed 
by the National Optical Astronomy Observatories, which is operated by 
AURA, Inc.\ under contract to the National Science Foundation.}, following 
the same procedure applied to the `central' field in Paper~I and to another 
field obtained in the same run \citep[west of NGC\,1399,][]{bas05}.  
This western field is not included in the present study since it
contains several galaxies with their own GCSs, which would severely
contaminate any search for NGC 1399 GCs (see Fig.~1). The seeing 
in the combined images was 1 arcsec on the $R$ frame and 1.3 arcsec on 
the $C$ frame.
 
The halo light of NGC\,1399 was 
subtracted by means of a ring median filter with an inner boundary 
of  1.3 arcsec and an outer boundary of 10.7 arcsec. This process 
also helped to correct a remaining flatfield structure present in the 
$C$ images that was not possible to remove even with master-flats. 
The sensitivity variations in the final combined images is below 0.3 percent.

The photometry was done with DAOPHOT within IRAF; DAOFIND on combined 
$C + R$ images was used for the first search. In the final $C$ and $R$ images 
and for each field, second-order variable PSFs were derived using about 100 
evenly  distributed stars per frame, which were fit to the sources 
through the ALLSTAR task. The estimated errors in the aperture 
corrections between the PSF radius (10\,pix) and a 15\,pix radius for the `east'  
$R$ and $C$ images were $\pm 0.009$ and $\pm 0.013$ mag, respectively, and 
for the `northeast' $R$ and $C$ images were $\pm 0.004$ and $\pm 0.008$ mag, 
respectively.

In order to perform the selection of point sources, the parameters $\chi$ and 
sharpness of the ALLSTAR task were used. Thus, over 5\,000 point sources 
were selected in each of the `east' and `northeast' fields, within the   
magnitude range $20 \la T_1 \la 25$.

The photometric calibration needed for the `east' and `northeast' fields 
has already been performed in Paper~I for the same run, so we will use those  
equations. We will not reproduce them here, so for further details please refer 
to Paper~I. As we intend to combine observations from the `central' field 
with the two adjoining ones, it is essential to preserve the homogeneity 
in the photometry. Thanks to the point sources located in the regions 
where the central field overlaps with the other two, we can compare the 
magnitudes (after calibration of each field to the standard system) and 
refer all the magnitudes to the `central' field system, 
particularly in this case where differences are likely to arise due 
to the large size of the images involved and remaining sensitivity differences 
between the eight chips of the MOSAIC fields. 
In this way, corrections (adjacent minus central field magnitude) 
$\delta R = 0.09 $ and $\delta C = 0.27 $ were  
applied to the `east' field point sources to refer its magnitudes to 
the `central' one, and $\delta R = 0.05$ and $\delta C = 0.14$ to the 
`northeast' field, with the same purpose. The number of overlapping 
objects used to calculate those corrections are: 189 objects in $C$ and 
489 in $R$ between the `central' and `east' fields, and 61 objects in both 
$C$ and $R$ between the `central' and `northeast' fields.  

We select as GC candidates those point sources with colours
$0.8 \leq (C-T_1) \leq 2.3$ and magnitudes $20 \leq T_1 \leq 23$. The 
adopted limiting magnitude $T_1 = 23$ is bright enough so that 
completeness corrections should only be a few percent and will not be 
necessary for the aim of 
this study. The separation between blue and red GC candidates will 
be taken as $(C-T_1) = 1.55$ following Paper~I. 
  
The selection of GC candidates should be corrected for contamination 
by the background. For this purpose, a background field located 3\fdg5 
northeast from NGC\,1399 was used, as in Paper~I. In this field, the 
projected density of point sources within the colour and magnitude ranges 
corresponding to blue GCs is $0.26 \pm 0.01$~objects\,arcmin$^{-2}$  
and the one corresponding to red GCs $0.10 \pm 0.01$~objects\,arcmin$^{-2}$.
In the rest of this paper, zero reddening has been assumed \citep{ost98}.

\section{Globular cluster colour distribution}

Figure~2 presents the colour-magnitude diagram for point sources in the whole 
field under study, where we have taken the average for GCs observed on more 
than one field. The blue and red GC candidates show up within the stated 
ranges of magnitudes and colours. This diagram looks very similar to the 
one obtained in Paper~I (see their Fig.~3). 

With the aim of analysing the colour distribution of the GC candidates at 
different projected distances from the NGC\,1399 centre, we plot the 
corresponding colour distributions in Fig.~3, smoothed with a 
Gaussian kernel with a dispersion 0.09 mag (the size of the histograms 
bins), for four different radial ranges: 1.5--9, 9--20, 20--35, 
and 35--52 arcmin. We choose an outer limit of 52 arcmin, as this is the 
largest distance from NGC\,1399 we can reach with a reliable areal coverage 
for estimating densities. These colour distributions were computed after excluding 
the GC candidates seen in projection close to other galaxies present in the 
field. In a conservative way we excluded the circles with radii: 
r = 1000\,pix (4.5 arcmin) around NGC\,1387, r = 600\,pix (2.7 arcmin) 
around NGC\,1404, and r = 2000\,pix (9 arcmin) around NGC\,1427 
(NGC\,1427 is on the eastern border of the `east' field, see Fig.~1). 
The raw colour distributions and the distributions 
with statistical background subtraction are displayed together with 
the histograms of the background-corrected data and the corresponding 
background colour distributions. The results from the fits of the sum of 
two Gaussians performed on the colour histograms are given in Table~1; 
the same results are obtained, within the uncertainties, if the fits 
are performed directly on the smoothed colour distributions instead 
of the histograms.  

\begin{figure*}
\begin{minipage}[c]{\hsize}
\centering
\includegraphics[width=0.47\hsize]{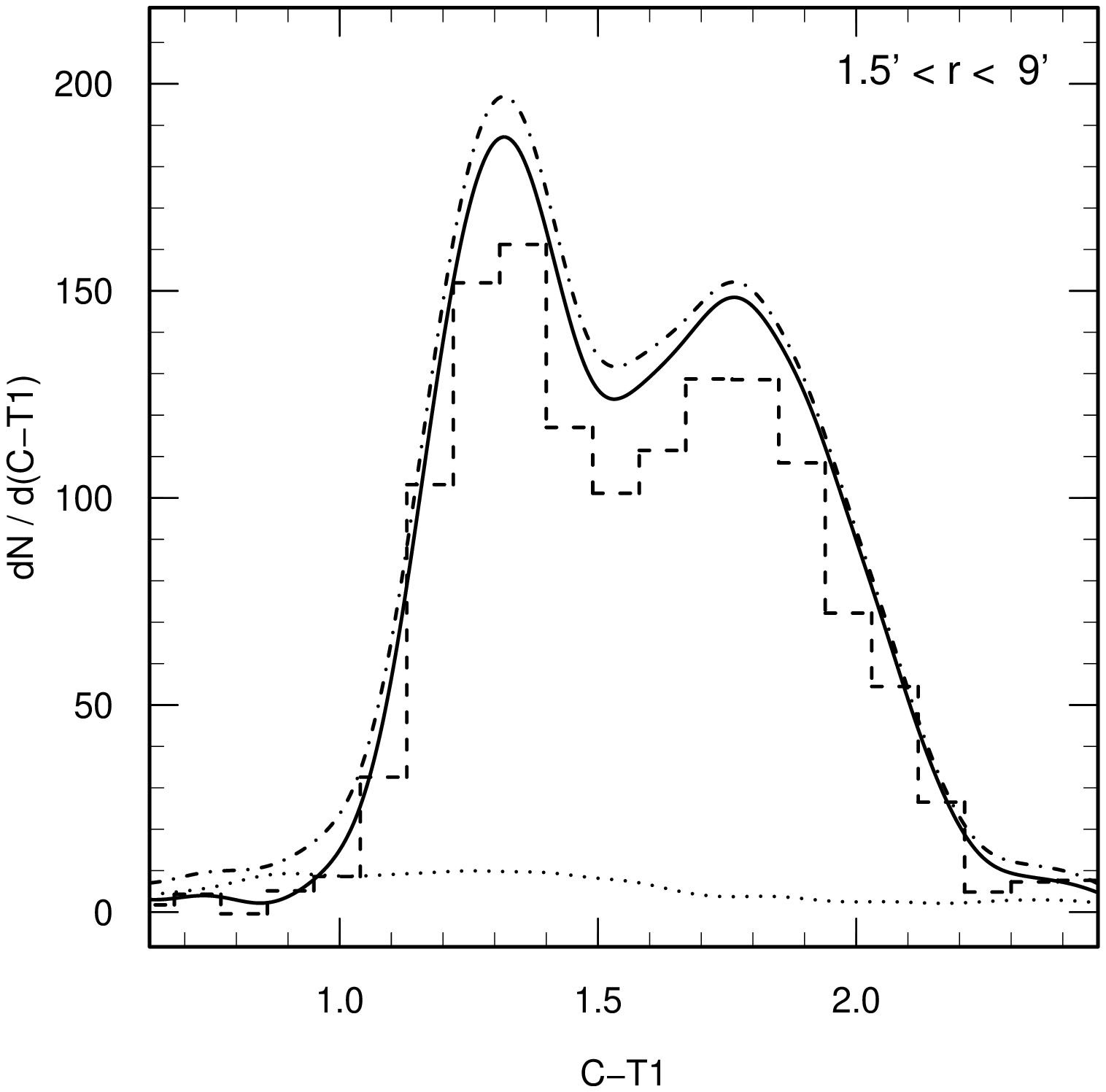}
\includegraphics[width=0.47\hsize]{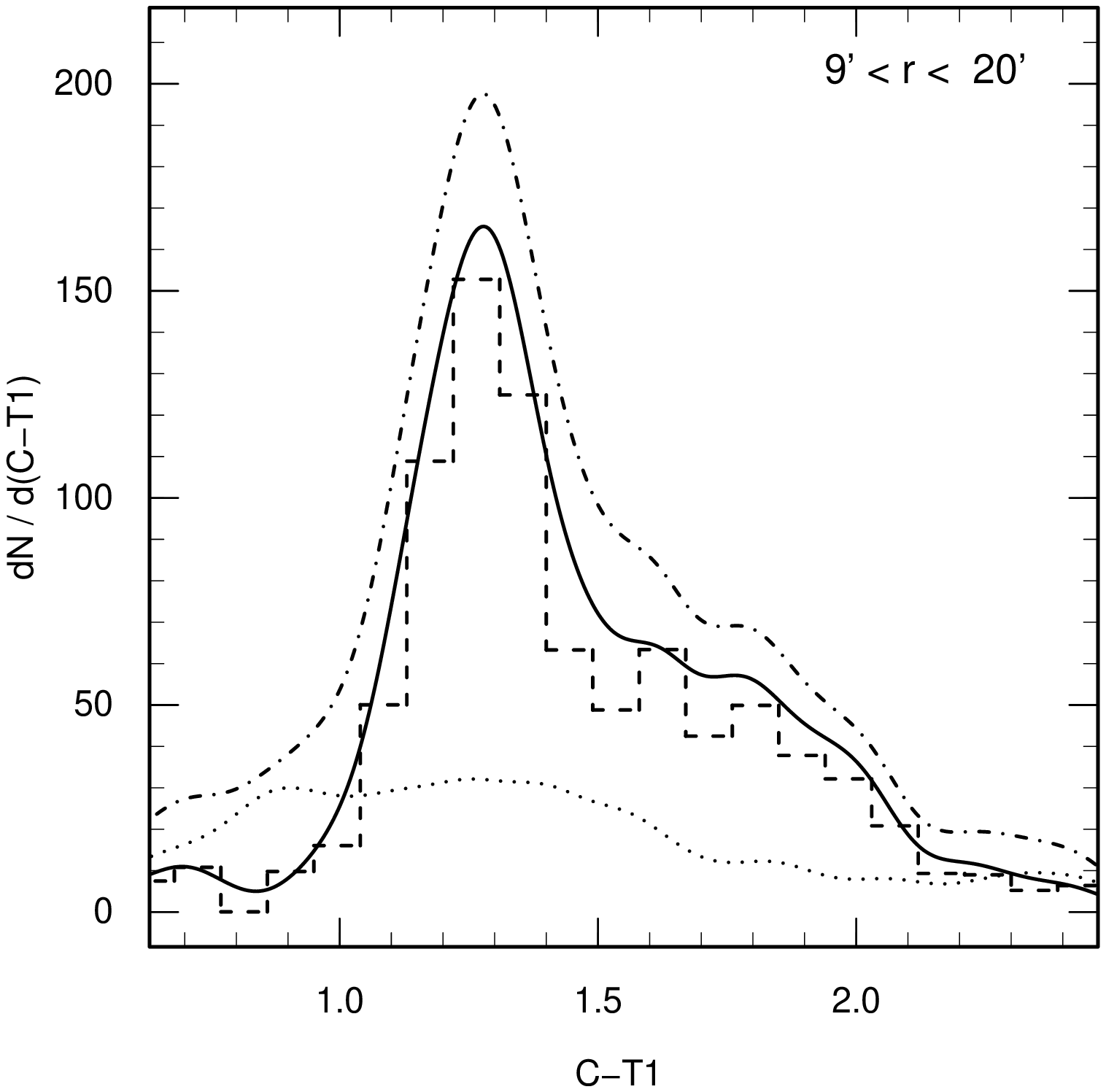}
\end{minipage}
\begin{minipage}[c]{\hsize}
\centering
\includegraphics[width=0.47\hsize]{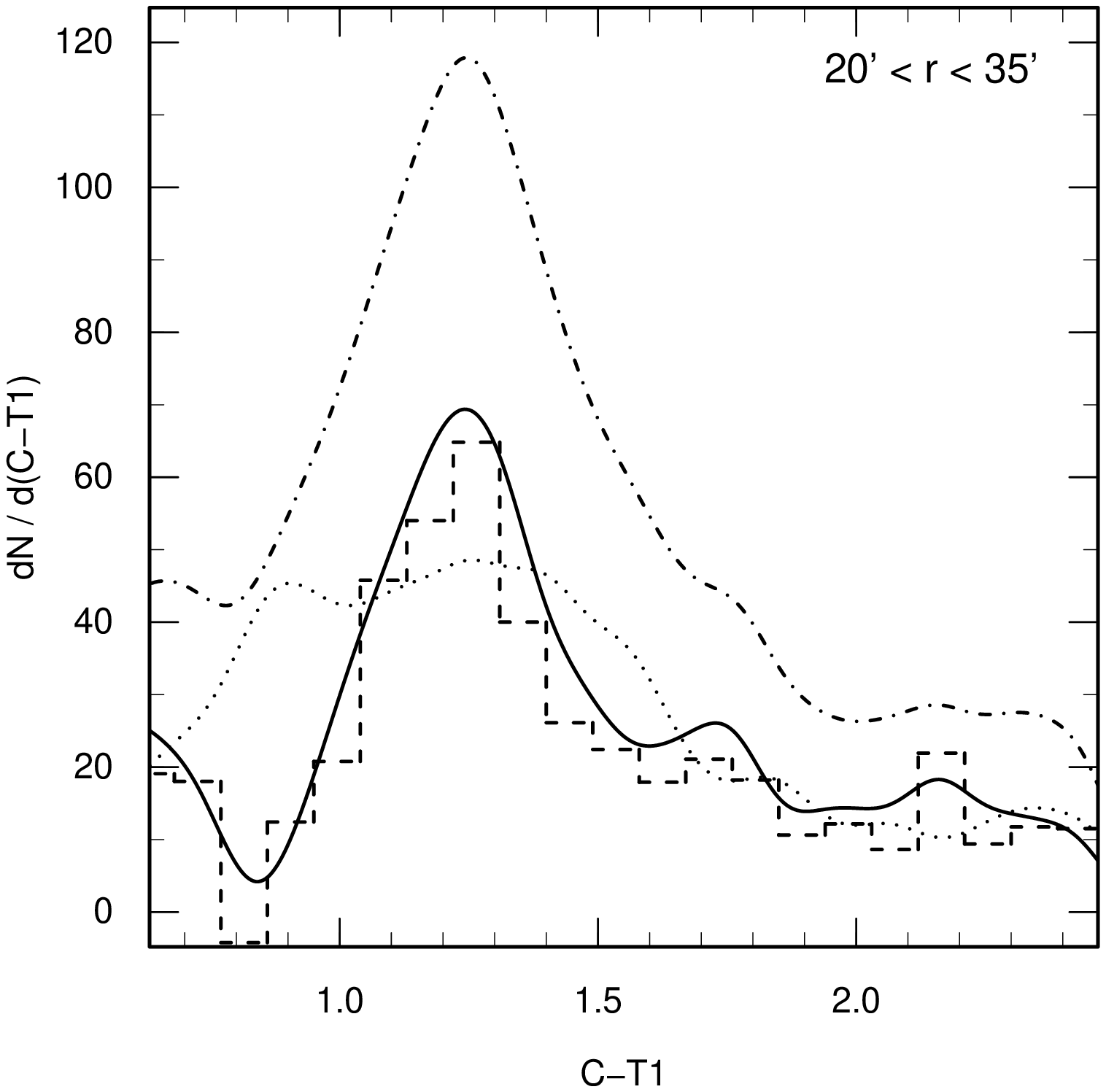}
\includegraphics[width=0.47\hsize]{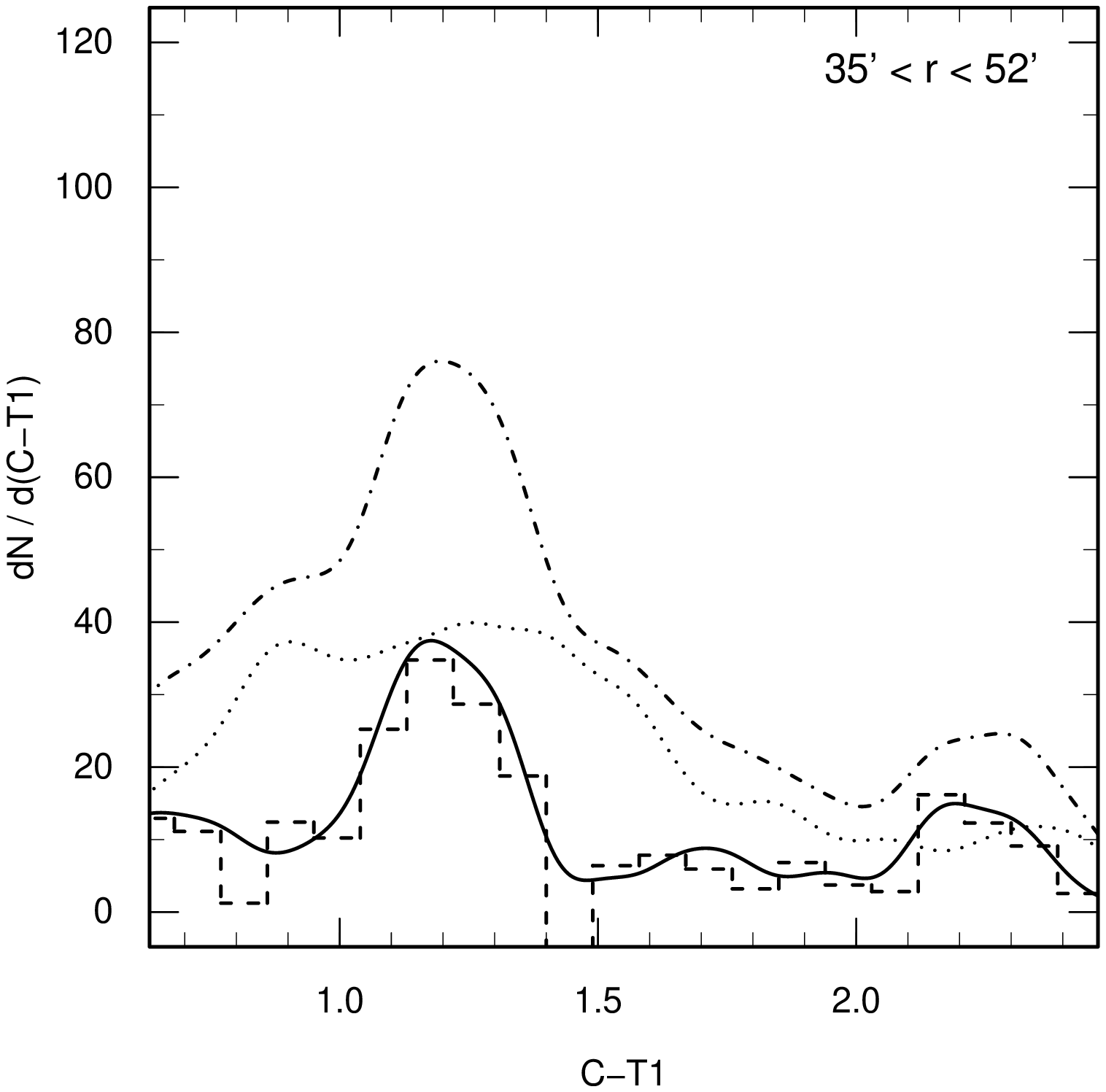}
\end{minipage}
\caption{Colour distribution functions for the GC candidates around NGC\,1399, 
for four different radial ranges indicated in the upper right corner within each 
panel. Dash-dotted and solid lines show the raw and background-corrected 
colour distributions, respectively (see text). The histograms of the 
background-corrected data are plotted with dashed lines, and the 
background colour distributions with dotted ones.}
\end{figure*}

\begin{table}
\centering
\caption{Results of the one or two Gaussian fit to the background-corrected 
colour distributions of the identified blue ($0.8 \leq (C-T_1) \leq 1.55$) 
and red ($1.55 \leq (C-T_1) \leq 2.3$) GC candidates with $20 \leq T_1 \leq 23$.}
\begin{tabular}{@{ } l @{~~~} c @{~~~~~} c @{~~~~~} c @{~~~~~} c @{ }}
\hline\hline
\noalign{\smallskip}
 r [arcmin] & $(C-T_1)_{{\rm blue}}$ & $\sigma_{{\rm blue}}$ & $(C-T_1)_{{\rm red}}$ & $\sigma_{{\rm red}}$\\
\noalign{\smallskip}
\hline
1.5 -- 9 & $1.29 \pm 0.01$ & $0.12 \pm 0.01$ & $1.76 \pm 0.01$ & $0.19 \pm 0.03$\\
9 -- 20  & $1.26 \pm 0.01$ & $0.10 \pm 0.01$ & $1.59 \pm 0.04$ & $0.10 \pm 0.01$\\
20 -- 35 & $1.23 \pm 0.02$ & $0.12 \pm 0.02$ &  --   &  --  \\
35 -- 52 & $1.18 \pm 0.03$ & $0.14 \pm 0.03$ &  --   &  --   \\
\hline
\end{tabular}
\end{table}

The innermost colour distribution (1.5 $<$ r $<$ 9 arcmin) is clearly bimodal 
with $C - T_1$ colours for the blue and red peaks in agreement, within the 
errors, with the values obtained in Paper~I for the 1.8--4.5 arcmin range.  
In the following sample (9 $<$ r $<$ 20 arcmin) the blue peak is slightly bluer 
than the inner one while the red GCs show a larger spread in colour, and are 
also bluer than their inner counterparts. In the third group 
(20 $<$ r $<$ 35 arcmin), the mean colour of the blue peak is again bluer, 
close to the limit of the errors, than the two inner samples, and the 
red GCs show a distribution that is marginally in excess of the background 
level; it is not possible to fit a Gaussian to the colour histogram 
of the red GCs in this case. 
Finally, even in the outermost sample (35 $<$ r $<$ 52 arcmin), the blue peak 
can be clearly identified and its position shows the same tendency; that is, 
it is bluer than all the previously determined ones, while the red GCs are 
absent, the colour distribution being indistinguishable from the background 
in their expected colour range. 

From these colour distributions it can be inferred that the mean colour 
of the blue peak gets bluer with increasing radius through 
the radial extent considered for computing the colour distributions, 
that is, from 1.5 up to 52 arcmin. If we assume that the blue peak colours 
correspond to the centre of each radial bin (Table~1) and transform the 
colours into metallicities by means of the relation given by \citet{har02}, 
we obtain a metallicity gradient for the blue GCs of  
$\Delta ~[Fe/H] ~/~ \Delta ~log(r) = - 0.28 \pm 0.06$ (r in kiloparsecs). 
The same effect has been found in the GCS around NGC\,1427 
\citep[ see their Fig. 14]{for01} where a colour gradient is present 
in the blue GC population; it agrees with our value within the 
(rather large) errors.   
We also show that the red GCs are present up to a shorter projected 
distance from the centre of NGC\,1399 than the blue ones, which 
supports the assumption that the blues have a shallower radial-density 
profile. The red GCs also show a significant 
gradient from the first to the second radial bins, although the peak in 
the second bin is harder to detect. 

\begin{figure}
\resizebox{\hsize}{!}{\includegraphics{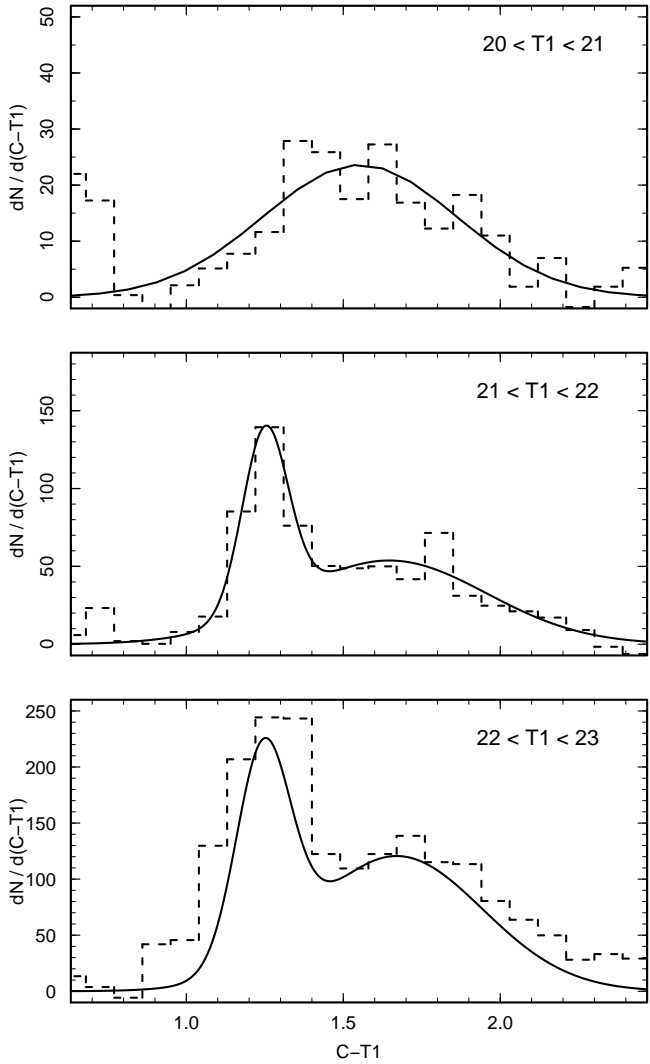}}
\caption{Colour-distribution functions for the GC candidates around NGC\,1399, 
for three different $T_{1}$ magnitude ranges indicated in the upper right 
corner within each panel. The histograms of the background-corrected 
colour distributions are plotted with dashed lines and the Gaussian fits 
with solid lines.}
\end{figure}

In order to confirm that out results do not rely on the chosen 
background colour distribution, we re-calculated the colour 
distribution functions for the GC candidates taking a limiting 
magnitude $T_1 = 22$, instead of $ T_1 = 23$ as considered in the rest 
of this paper. The resulting colour distributions (not displayed) show 
that taking a limiting magnitude $ T_1 = 22$ for the raw GC and the 
background colour distributions does not change the shapes 
of the final colour distributions (background corrected) 
in the three inner radial bins. Even the mean colours of the blue and red 
peaks agree, within the errors, with the ones calculated with limiting 
magnitude $ T_1 = 23$. 
The only difference appears in the outer bin (35 $<$ r $<$ 52 arcmin) because 
there are almost no (blue) GC candidates left adopting a brighter limiting 
magnitude. In this latter bin, the raw colour distribution looks 
very similar to that 
of the background, which can be considered additional evidence that 
the colour distribution of the chosen background field is robust.

The colour distributions for three different magnitude ranges and without 
any radial selection are displayed in Fig.~4. It is interesting to note 
that the GCs belonging to the brightest bin (upper panel) do not seem to 
present a bimodal distribution, as already suggested by \citet{ost98} 
and in Paper~I, but a unimodal, very broad  distribution with a mean colour 
$C - T_{1} = 1.55 
\pm 0.06$. The two fainter brightness bins show bimodal distributions 
with blue and red peak positions in excellent agreement: for the 
$21 < T_{1} < 22$ interval the blue/red peak colours are 
$C - T_{1} = 1.25 \pm 0.01 / 1.65 \pm 0.05$, while for the $22 < T_{1} < 23$ 
range we obtain $C - T_{1} = 1.24 \pm 0.01 / 1.67 \pm 0.05$.  
 
\section{Globular-clusters' spatial distribution}

All blue and red GC candidates identified in the three MOSAIC fields are 
shown in Fig.~5. It can be clearly seen that the red GCs (lower panel) 
present a more concentrated projected distribution than the blue ones 
(upper panel), with respect to the galaxies NGC\,1399, NGC\,1387,  
and NGC\,1404.   

\begin{figure}
\resizebox{\hsize}{!}{\includegraphics{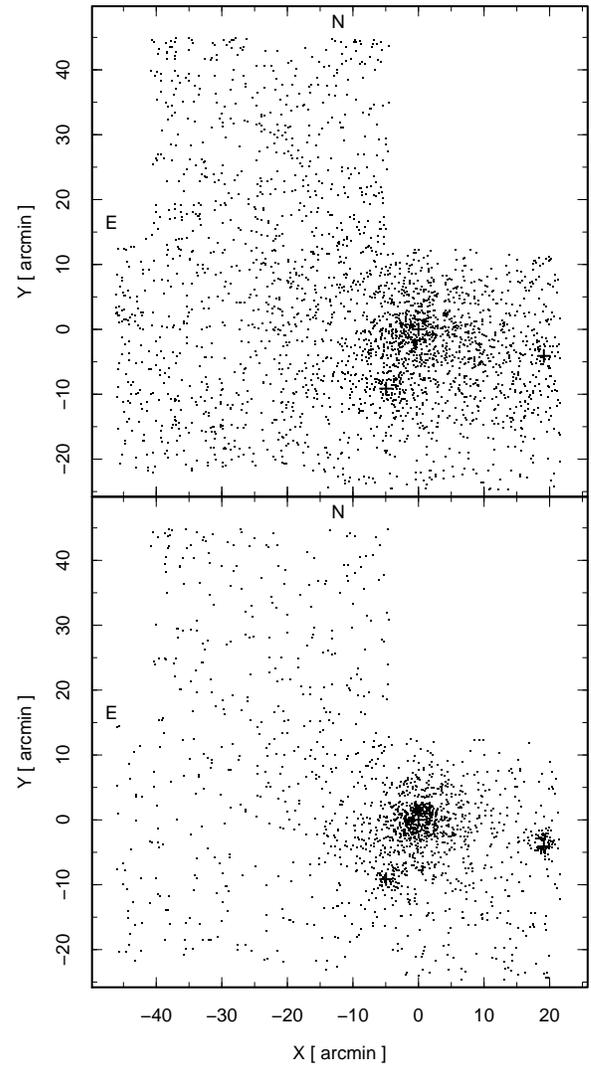}}
\caption{Projected spatial distribution of blue (upper panel) and red (lower 
panel) GC candidates on the MOSAIC fields. The crosses identify, 
from east to west, the centres of the galaxies NGC\,1404, 
NGC\,1399 (at 0,0), and NGC\,1387.}
\end{figure}

\subsection{Radial distribution}

Figures~6 and 7 show the radial number-density profiles of the blue and 
red GCs around NGC\,1399, respectively. The profiles were measured excluding 
the GC candidates located in the vicinity of NGC\,1387, NGC\,1404, and 
NGC\,1427 as already explained in Sec. 3. The profiles were corrected 
for contamination by the background using the field located 3.5 deg 
northeast of NGC\,1399, and the Poisson's errors include the errors of 
the raw counts and of the background. Both figures show power-law fits 
to these background-corrected density profiles in the upper panels and 
r$^{1/4}$ fits in the lower ones. Table~2 shows the background-corrected 
radial number densities for blue and red GCs and the fraction of the rings  
sampled in each case.

\begin{figure}
\resizebox{\hsize}{!}{\includegraphics{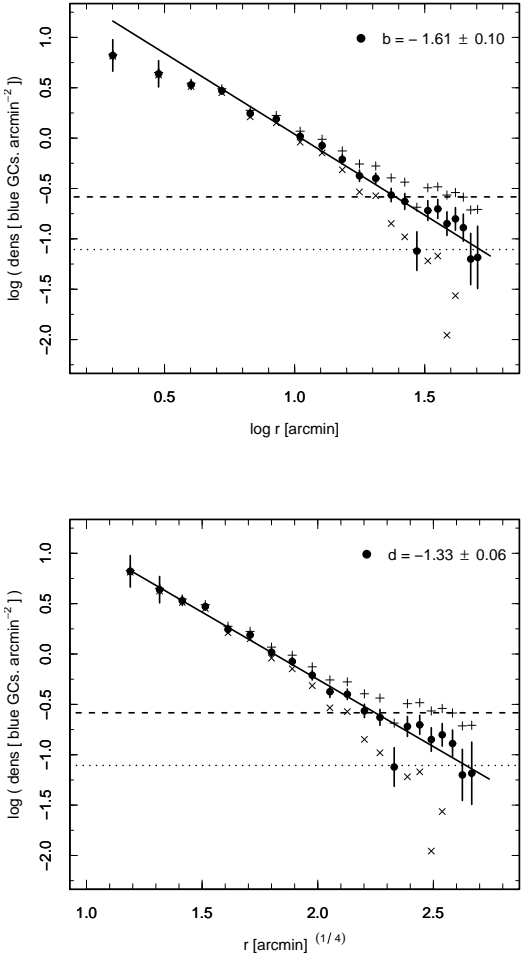}}
\caption{Radial density profiles for blue GC candidates in the field of 
NGC\,1399 showing power law (upper panel) and $r^{1/4}$ fits 
(lower panel) to the data. Filled circles show the background-corrected 
distribution of blue GC candidates, while plus signs and crosses show the same 
distribution considering a background 50 per cent higher and 50 per cent 
lower, respectively, than the original one. The dashed lines correspond to 
the background projected density and the dotted lines to 30 per cent of 
the background density. The coefficients in the upper right corners give 
the slope of the corresponding fits.  
}
\end{figure}

\begin{figure}
\resizebox{\hsize}{!}{\includegraphics{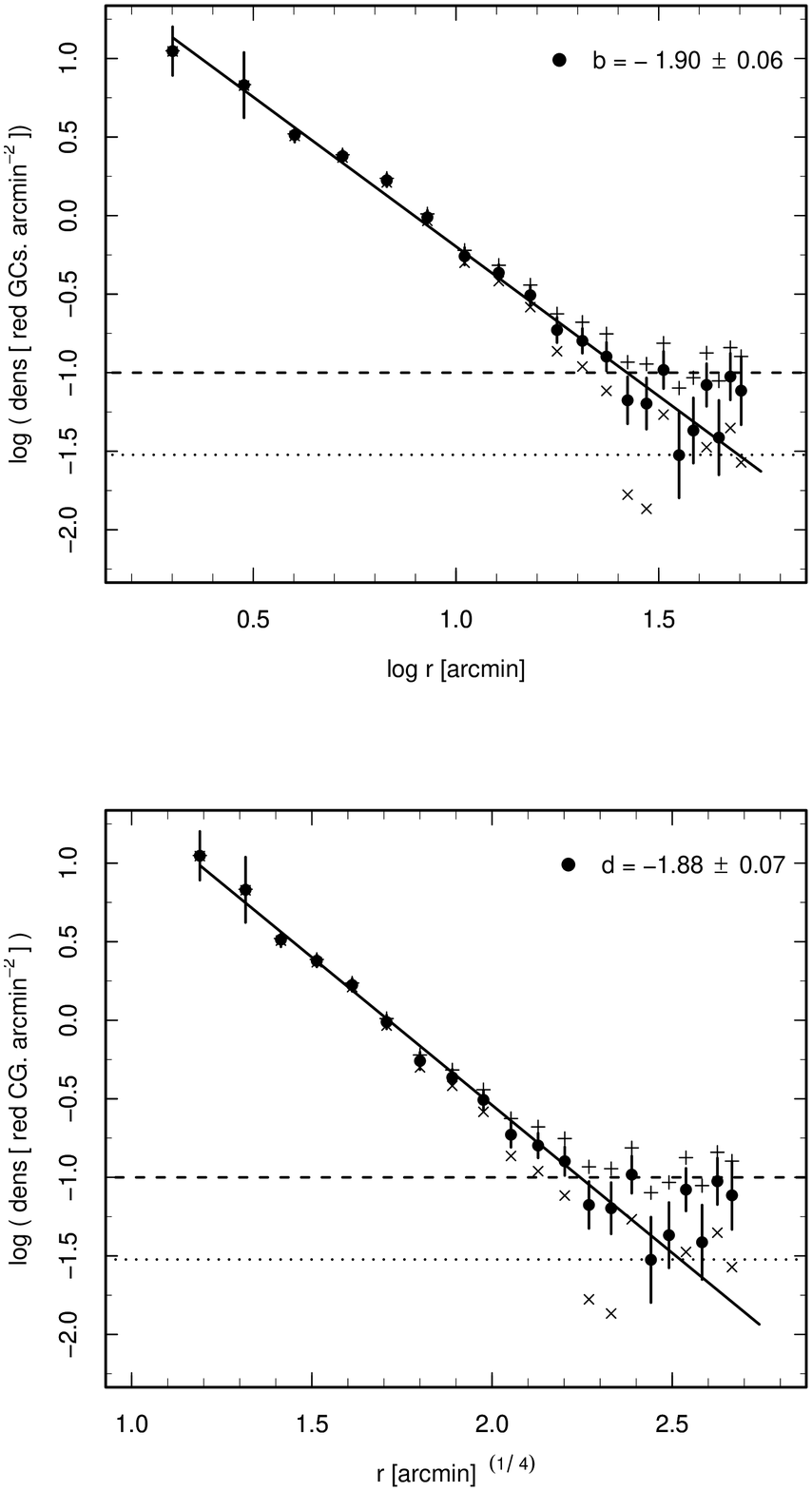}}
\caption{As in Fig. 6 but for red GCs in the field of NGC\,1399. 
  }
\end{figure}

\begin{table}
\centering
\caption{Radial number densities $\delta$ (background-corrected$^{\mathrm{a}}$) 
of blue and red GC candidates in NGC\,1399, in GC\,arcmin$^{-2}$.}
\begin{tabular}{lccc}
\hline\hline
\noalign{\smallskip}
r~[arcmin] & $\delta_{{\rm blueGC}}$ & $\delta_{{\rm redGC}}$ & \% sampled ring\\
\noalign{\smallskip}
\hline
1.5 -- 2.5 & $6.62 \pm 0.70$ & $11.14 \pm 0.89$ &  100\\
2.5 -- 3.5 & $4.35 \pm 0.50$ & $6.77 \pm 0.61$ &     100\\
3.5 -- 4.5 & $3.38 \pm 0.38$ & $3.26 \pm 0.37$ &     100\\
4.5 -- 6   & $2.96 \pm 0.25$ & $2.39 \pm 0.22$ &     100\\
6~ -- 7.5   & $1.76 \pm 0.18$ & $1.68 \pm 0.16$ &    100\\
7.5 -- 9.5 & $1.55 \pm 0.14$ &  $0.98 \pm 0.10$ &    93\\
9.5 -- 11.5 & $ 1.04 \pm 0.11$ & $0.55 \pm 0.08$ &     88\\  
11.5 -- 14  & $0.84 \pm 0.08$ & $0.43 \pm 0.05$ &     89\\
14~ -- 16.5  & $0.62 \pm 0.07$ & $0.31 \pm 0.05$ &     77\\
16.5 -- 19 & $0.42 \pm 0.06$ & $0.19 \pm 0.04$ &     75\\
19 -- 22   & $0.40 \pm 0.05$ &  $0.16 \pm 0.03$ &    72\\
22 -- 25   & $0.27 \pm 0.04$ & $0.13 \pm 0.03$ &     65\\
25 -- 28   & $0.24 \pm 0.05$ &  $0.07 \pm 0.03$ &    49\\
28 -- 31   & $0.08 \pm 0.04$ &  $0.06 \pm 0.03$ &    38\\
31 -- 34   & $0.19 \pm 0.05$ &  $0.10 \pm 0.03$ &    34\\
34 -- 37   & $0.20 \pm 0.05$ &  $0.03 \pm 0.02$ &    32\\
37 -- 40   & $0.14 \pm 0.04$ &  $0.04 \pm 0.03$ &    32\\
40 -- 43   & $0.16 \pm 0.05$ &  $0.08 \pm 0.03$ &    27\\
43 -- 46   & $0.13 \pm 0.05$ &   $0.04 \pm 0.03$ &   23\\
46 -- 49   & $0.06 \pm 0.05$ &  $0.09 \pm 0.04$ &    16\\
49 -- 52   & $0.06 \pm 0.06$ &   $0.08 \pm 0.05$ &    8\\
\hline
\end{tabular}
\begin{list}{}{}
\item[$^{\mathrm{a}}$] The projected densities of point sources in the 
background field within the colour and magnitude ranges corresponding 
to blue/red GCs are  0.26~/~0.10~objects\,arcmin$^{-2}$, respectively 
(limiting magnitude $T_1 = 23$).
\end{list}
\end{table}

The wide field covered in this study allowed us, for the first time, to perform 
an estimation of the total extent of the GCS around NGC\,1399. 
We adopted the galactocentric radius at which the areal density of blue GCs 
(corrected by background contamination) falls to 30 per cent of the 
background level as an indicative value for this extent, as it is the 
largest radial distance from NGC\,1399 at which we could separate the blue 
GCs from the background with our data. From the blue GC profile (Fig.~6) 
we obtained an extension of 45 arcmin that, at the Fornax distance, corresponds 
to a radial distance of 250 kpc. 
The same limit could not be obtained from the red GCs profile (Fig.~7) as 
the profile seemed to level out at a shorter radial distance, and there were  
fewer of them. It is at least worth noting that there are many red GCs 
lying out at large distances like 25 arcmin, that is, about 140 kpc. 

In order to compute the uncertainties involved in this determination due 
to errors in the background, independently of the pure Poissonian errors, 
we recalculated the profiles using `blue' and `red' backgrounds 50 per cent 
above and below the original ones (see Figs.~6 and 7). In this way, 
the outer limit for the blue GCS around NGC\,1399 is set to 45 $\pm$ 5 arcmin, 
that is, a range in radial extension from 220 to 275 kpc. 
This indicative value should be considered as a lower limit,  
as we cannot guarantee that it does not extend further out. 

The background-corrected density profile for blue GCs was first fit by 
a power law of the form $\delta =ar^{b}$ in the range 5 to 52 arcmin, 
excluding the outer regions, where there are very few candidates left,  
and the innermost region (r $<$ 5 arcmin) that cannot be fit 
with the same power law (see Fig.~6, upper panel). 
                                 
The density profile for red GCs was also fit by a power law but 
over a shorter range r $<$ 35 arcmin, which is the projected angular 
distance at which the profile levels out (Fig.~7, upper panel); in this 
case, the innermost region did not show any discrepancy with the power-law fit.  
The slope obtained for blue GC candidates was $b = -1.61 \pm 0.10$ 
(rms = 0.13), while for red candidates $b = -1.90 \pm 0.06$ (rms = 0.09). 
Thus, the red GCs around NGC\,1399 show a steeper profile than the blues, 
even at large radial distances. 

In Paper~I, the blue clusters also showed a shallower density profile than
the red clusters within 7 arcmin ($b = -0.8 \pm 0.17$ and $b = -1.64 \pm 0.10$ 
for blues and reds, respectively), but could not be distinguished outside 
this radius. The reason for this discrepancy with Paper~I is the steeper 
slope found here for the red clusters. 

It is interesting to note that both density profiles, for blue and red 
clusters, can also be fit reasonably well by r$^{1/4}$ laws: $\log\,\delta = 
c + dr^{1/4}$ (Figs.~6 and 7, lower panels). In this case, 
the blue GCs profile can be fit even in the innermost region, and 
we obtain $d = -1.33 \pm 0.06$ (rms = 0.12) for the blue candidates 
within 52 arcmin. 
The reds' density profile was again fit up to r $<$ 35 arcmin, as in the 
power-law fit, for the same reason stated above; a slope 
$d = -1.88 \pm 0.07$ (rms = 0.10) is obtained for the red candidates. 

\subsubsection{Comparison of the radial density profile of blue clusters  
with a NFW profile}

The volumetric density function $\rho(r)$ adopted by \citet{ric04} 
for their dark matter halo  
\begin{equation}
      \rho(r) = \frac{\rho_s}{(r/r_s)^\zeta \, (1+r/r_s)^{3-\zeta}} \,,
\end{equation}
\noindent where $\rho_s$ and $r_s$ are the characteristic density and radius, 
respectively, corresponds to the NFW profiles \citep{nav97} in the case 
$\zeta = 1$. 

The linear least-square regression of the areal density for the blue globular 
clusters as a function of $r^{1/4}$ (depicted in Fig.~6, lower panel) has led to: 
$\log\,\delta~=~2.42~-~1.33~r^{1/4} $, with an rms = 0.12 (including all 
the data points). This fit is practically equivalent to a projected 
NFW profile with a scale length $r_s$ = 6.25 arcmin ($\zeta = 1.0$)  
that, as shown in \citet{for05}, provides a good approximation 
of the blue clusters inside a galactocentric radius of 16.7 arcmin. 
It is interesting to note that such a NFW profile is indistinguishable from the 
NFW profile (for $\zeta = 1.0$) obtained for the dark halo of NGC\,1399 in 
the kinematic and dynamical study of its GCS performed by \citet{ric04}.    
Fig.~8, on the other side, shows that the same regression holds in the 
$\log\,\delta$ vs. $\log\,r$ plane, implying that a single power would be 
adequate and that multiple power-law fits do not seem required.

\begin{figure}
\resizebox{\hsize}{!}{\includegraphics{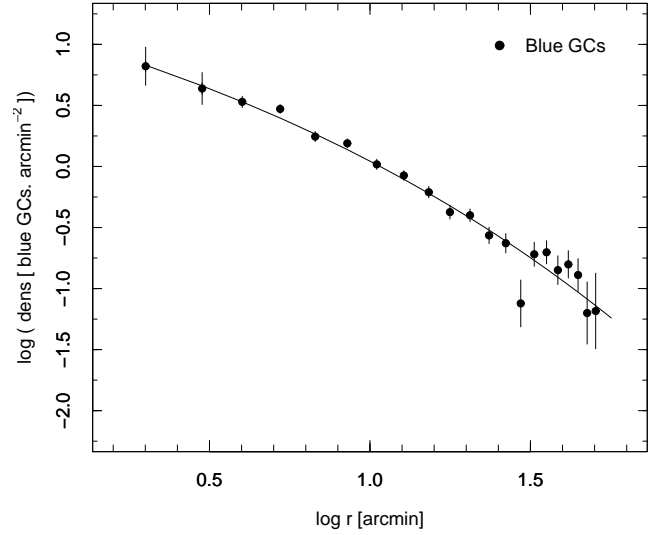}}
\caption{Linear least-square regression of the areal density as a function 
of $r^{1/4}$ for the blue globular clusters, displayed in the plane 
log~(density) vs. log r.}
\end{figure}

The linear least-square regression of the areal density for the red globular 
clusters as a function of $r^{1/4}$ (shown in Fig.~7, lower panel) yielded: 
$\log\,\delta~=~3.22~-~1.88~r^{1/4}$, with rms = 0.10. This solution, 
as mentioned above, includes only the points within r $<$ 35 arcmin (the 15 
innermost data points) due to the increase of the profile noise outwards. 
A slightly better result, in terms of the rms, is obtained with a projected 
generalized NFW profile and after adopting the same parameters as in 
\citet{for05}, i.e., $\zeta = 1.5$ and a scale length $r_s$ =0.5 arcmin  
(rms = 0.08). As noted in that paper, this profile, in turn, 
cannot be distinguished from a Hubble profile with a core radius of about 
one arcmin \citep[see][]{for97b}.
  
\subsection{Azimuthal distribution}

We analysed the azimuthal distribution of blue GCs around NGC\,1399, which is 
depicted in Fig.~9, to search for any dependence of GC surface density on  
the position of neighbouring galaxies. The smoothed, projected distribution 
was obtained through a Gaussian kernel with a dispersion of 54 arcsec. Curves 
of constant numerical density have been overlaid on the image.

\begin{figure}
\resizebox{\hsize}{!}{\includegraphics{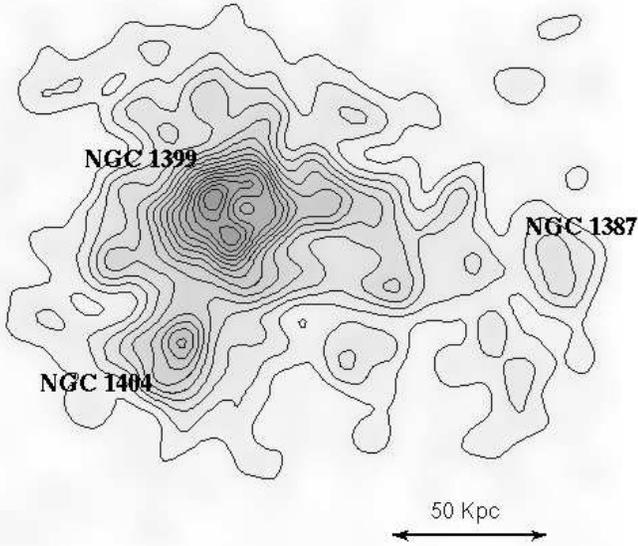}}
\caption{Smoothed, projected distribution of blue GC candidates around NGC\,1399. 
Solid lines correspond to curves with constant numerical density. North is up 
and East to the left.}
\end{figure}

Two tails can be seen in the smoothed blue GC projected distribution, 
one towards NGC\,1404 and another towards NGC\,1387. As the angular separation 
between NGC\,1399 and NGC\,1404 is just 10 arcmin (a projected distance of 
55 kpc at the Fornax distance), it is not clear if the `tail' in that direction 
is showing some kind of interaction between the two galaxies that can be traced 
by the GCs or if it is just the superposition of the NGC\,1399 and NGC\,1404 GCSs. 
                                        
On the other hand, the angular distance between NGC\,1399 and NGC\,1387 is 
about 19 arcmin, while the radial extension of the GCS around NGC\,1387 is 
estimated as 3.1 arcmin \citep{bas05}. In this case, 
the tail of the smoothed, projected distribution of blue GCs is pointing to an 
excess of such globulars in that direction, i.e. at a position angle 
$\approx 260 \deg$ (measured from N to E with respect to NGC\,1399). 
We are then led to assume that some kind of interaction may be under way 
between both galaxies, and that blue GCs are tracing it. 
We discuss this point further below.

\section{Summary and discussion}

This large-scale study of the NGC\,1399 GCS allowed us to analyse the 
colour and spatial-projected distribution up to a projected distance of 
52 arcmin (about 290 kpc), i.e. up to a distance that has not been 
reached before with this kind of work.

The colour distribution at different galactocentric distances shows that the 
bimodality, present in the inner samples, gradually turns into a single blue 
population in the outer bins, in agreement with the general idea that red GCs 
have a steeper radial density profile than the blues, which is confirmed 
by the projected GC areal density profiles shown in Figs.~6 and 7. Besides, 
the position of the blue peak moves to bluer colours at larger galactocentric 
radii, a gradient that has also been detected in the NGC\,1427 GCS by 
\citet{for01}.  

In addition, the colour distribution at different magnitude ranges 
shows bimodal distributions for the intermediate and faint samples, but it 
seems to be unimodal at the brighter bin ($20 < T_1 < 21$), in agreement 
with Paper~I. It is interesting to note that a similar behaviour has also 
been detected in the GCSs of several brightest cluster galaxies by 
\citet{har05} and of the Virgo giant ellipticals M\,87 and NGC\,4649 
by \citet{str05}. 
On the other side, the colour distribution 
at the bright magnitude range $20 < T_1 < 22$ of the GCs around NGC\,4636 is 
not unimodal but bimodal with a very similar number of blue and red clusters 
\citep{dir05}.  

With respect to the radial projected distribution, Figs.~6 and 7 show that the 
outer regions of NGC\,1399 are in fact dominated by an extension of its inner 
globular cluster populations out to at least 45 arcmin (or 250 kpc) 
in galactocentric radius, a range comparable to the Fornax cluster 
core radius \citep[about 40 arcmin,][]{fer89}.
There is remarkable agreement between the blue GC radial density profile and 
the NFW profile \citep{nav97} for the dark-matter distribution obtained by 
\citet{ric04} by means of their kinematical study of the NGC\,1399 GCS. 

Two tails show up in the (smoothed) azimuthal projected distribution of blue 
GCs around NGC\,1399. One of them, towards NGC\,1404, is likely to arise 
as an overlapping of both GCSs, although some authors point to an interaction 
process between both galaxies that leads to a stripping of GCs from 
NGC\,1404 \citep{for97a,bek03}. 
However, the tail towards NGC\,1387 cannot be explained as just GCSs 
overlapping due to the larger projected distance between 
NGC\,1399 and this galaxy. This evidence of an excess of blue GCs in that 
particular direction, as well as the low number of blue GCs present around 
NGC\,1387 as compared to the red ones \citep{bas05}, support 
the idea that this might be a case of tidal stripping 
\citep[see, for instance,][]{for82,muz87}, a process through which part of 
the GCs from NGC\,1387 might be being captured by the (much more massive) 
NGC\,1399. Or during that process, they just remain within the potential 
well of the cluster without being bound to any individual galaxy 
--the intracluster GCs \citep{bas03}. Such scenarios  
have already been suggested by \citet{for97a}.   
It is also expected that this process will primarily affect the 
blue globulars as they represent the more extended GC sub-population. 
Radial velocities will probably help in confirm this assumption.
 
\begin{acknowledgements}

We wish to thank the referee for comments that helped to
improve the present paper.
This work was funded with grants from the Consejo Nacional de Investigaciones
Cient\'{\i}ficas y T\'ecnicas de la Rep\'ublica Argentina, Agencia Nacional
de Promoci\'on Cient\'{\i}fica Tecnol\'ogica, and Universidad Nacional de La
Plata (Argentina). D.G. and T.R. gratefully acknowledge support from the 
Chilean {\sl Centro de Astrof\'\i sica} FONDAP No. 15010003. LPB is grateful 
to the Astronomy Group at the Universidad de Concepci\'on 
for financial support and warm hospitality. Y.S. gratefully 
acknowledges support from a German Science Foundation 
Grant (DFG--Projekt HI-855/2).

\end{acknowledgements}

\end{document}